\documentclass[useAMS,usenatbib,usegraphicx]{mn2e}
\usepackage{amsmath}

\def\OI{[\mbox{O\,{\sc i}}]} 
\def\OIII{[\mbox{O\,{\sc iii}}]}
\def\NII{[\mbox{N\,{\sc ii}}]}
\def\SII{[\mbox{S\,{\sc ii}}]}
\def\Ha{\mbox{H$\alpha$}}
\def\Hb{\mbox{H$\beta$}}
\newcommand{\gtsim}{\mbox{{\raisebox{-0.4ex}{$\stackrel{\star}
{{\phantom{o}}} $}}}}
\title[Spectral types of southern ULIRGs]
{Optical spectral classification of southern ultraluminous infrared galaxies}
\author[J. C. Lee et al.]
{Jong Chul Lee$^{1}$\thanks{Visiting Astronomer, Cerro Tololo Inter-American Observatory, National Optical Astronomy Observatory, 
which are operated by the Association of Universities for Research in Astronomy, 
under contract with the National Science Foundation.}\thanks{E-mail: jclee@astro.snu.ac.kr},
Ho Seong Hwang$^{2}$\gtsim,
Myung Gyoon Lee$^{1}$\thanks{E-mail: mglee@astro.snu.ac.kr},
Minjin Kim$^{3}$,
\newauthor and Sang Chul Kim$^{4}$\gtsim \vspace{0.2cm}\\
$^{1}$Astronomy Program, Department of Physics and Astronomy, Seoul National University, Seoul 151-742, Korea\\
$^{2}$CEA Saclay/Service D'Astrophysique, F-91191 Gif-sur-Yvette, France\\
$^{3}$National Radio Astronomy Observatory, 520 Edgemont Road, Charlottesville, VA, USA\\
$^{4}$Korea Astronomy and Space Science Institute, Daejeon 305-348, Korea}

\begin{document}

\date{Accepted 2011 January 27. Received 2011 January 26; in original form 2010 December 7}

\pagerange{\pageref{firstpage}--\pageref{lastpage}} \pubyear{2010}

\maketitle

\label{firstpage}

\begin{abstract}

We present a study of the optical spectral properties of 115 ultraluminous infrared galaxies
(ULIRGs) in the southern sky.
Using the optical spectra obtained at CTIO 4 m and provided by the 2dF Galaxy Redshift Survey and the 6dF Galaxy Survey,
we measure emission line widths and fluxes for spectral classification.
We determine the spectral types of ULIRGs with \Ha\ measurement using the standard diagnostic diagrams.
For ULIRGs without \Ha\ measurement, we determine their spectral types using the plane of
flux ratio between \OIII$\lambda5007$ and H$\beta$ versus \OIII\ line width based on our new empirical criterion.
This criterion is efficient to distinguish active galactic nuclei (AGNs) from non-AGN galaxies
with completeness and reliability of about 90 per cent.
The sample of 115 ULIRGs is found to consist of 8 broad-line AGNs, 49 narrow-line AGNs, and 58 non-AGNs.
The AGN fraction is on average 50 per cent and increases with infrared luminosity and {\it IRAS} 25$-$60 $\mu$m colour, 
consistent with previous studies.
The {\it IRAS} 25$-$60 $\mu$m colour distributions are significantly different between AGN and non-AGN ULIRGs,
while their {\it IRAS} 60$-$100 $\mu$m colour distributions are similar.
\end{abstract}


\section{Introduction}

Ultraluminous infrared galaxies (ULIRGs) with infrared luminosity at 8--1000 $\mu$m greater than 10$^{12}$ $L_{\odot}$ (Soifer et al. 1987)
are extremely energetic objects in the universe.
Although they contribute little to the infrared luminosity density in the local universe due to small numbers, 
they become cosmologically important at z~$>$~1 (e.g., Le Floc'h et al. 2005; Magnelli et al. 2009).
Their enormous infrared luminosity comes from dust heated by 
hot young stars (starburst), a supermassive black hole rapidly accreting matter (active galactic nucleus, AGN), or a mixture of these two
(see Sanders \& Mirabel 1996 and Lonsdale et al. 2006 for a review).
These starburst and/or AGN activities can be triggered by tidal interactions between galaxies and associated shocks 
(e.g., Bushouse 1987; Liu \& Kennicutt 1995; Barnes 2004).
In fact, numerous observational and theoretical studies suggested that ULIRGs are mergers of gas-rich disk galaxies
(e.g., Clements et al. 1996; Mihos \& Hernquist 1996; Veilleux et al. 2002; Younger et al. 2009; Hwang et al. 2010a),
and evolve into quasars (e.g., Sanders et al. 1988; Dasyra et al. 2006; Hopkins et al. 2006; Yuan et al. 2010)
or intermediate-mass elliptical galaxies (e.g., Genzel et al. 2001; Tacconi et al. 2002).

To better understand the origin and evolution of ULIRGs, it is essential to find out what their primary energy source is.
A standard spectroscopic method to distinguish between starburst and AGN is to use so-called BPT diagrams (Baldwin et al. 1981),
which are based on optical emission line ratios sensitive to the photoionization source.
This optical diagnostic was revised by Veilleux \& Osterbrock (1987)
and an alternative scheme was proposed by Kewley et al. (2006). %to separate starburst-AGN composite galaxies.
In heavily obscured galaxies like ULIRGs, additional observations at other wavelengths
are helpful to characterize their dominant energy source (e.g., X-ray: Franceschini et al. 2003, Teng et al. 2009;
infrared: Risaliti et al. 2006, Farrah et al. 2007, Imanishi et al. 2010; radio: Nagar et al. 2003, Sajina et al. 2008).
However, it is still difficult to determine the relative contribution of starburst and AGN within individual galaxies.

ULIRGs were discovered in large numbers by the {\it Infrared Astronomical Satellite} ({\it IRAS}; Neugebauer et al. 1984)
and the number of ULIRGs increased with the advent of wide-field galaxy redshift surveys
(Goto 2005; Pasquali et al. 2005; Cao et al. 2006; Hwang et al. 2007, 2010a; Hou et al. 2009).
However, previous studies based on optical spectra of the large sample are limited to the Sloan Digital Sky Survey
(SDSS; York et al. 2000), which mainly covers the northern hemisphere.
The optical spectral properties of ULIRGs in the southern hemisphere remain to be studied.

In this study, we present the analysis of optical spectra for about a hundred southern ULIRGs in the catalogue given by Hwang et al. (2007).
The structure of this paper is as follows.
The survey data and our observations are described in Section 2.
Section 3 explains the procedures to analyse these data and to classify the ULIRGs.
In Section 4, the results of our study are discussed, and are compared with those of previous studies.
We summarize and conclude in Section 5.
Throughout, we adopt $H_{0}$ = 70 km s$^{-1}$ Mpc$^{-1}$ and a flat $\Lambda$CDM cosmology with density parameters $\Omega_{M}$ = 0.3
and $\Omega_{\Lambda}$ = 0.7.

\section{Observations and data}

Hwang et al. (2007) identified 324 ULIRGs by
cross-correlating the {\it IRAS} Faint Source Catalogue Version 2 (Moshir et al. 1992)
with the spectroscopic catalogues of galaxies in the Fourth Data Release of SDSS (Adelman-McCarthy et al. 2006.),
the Final Data Release of the 2dF Galaxy Redshift Survey (2dFGRS; Colless et al. 2001),
and the Second Data Release of the 6dF Galaxy Survey (6dFGS; Jones et al. 2004, 2005).
The spectral types of ULIRGs in the SDSS among them were determined in Hou et al. (2009).
In this study, we focus on 198 ULIRGs that were not covered by the SDSS but covered by the 2dFGRS and 6dFGS.
We used the optical spectra of these ULIRGs available online\footnote{For 2dFGRS, http://www.mso.anu.edu.au/2dFGRS/\\
$~~~$ For 6dFGS, http://www-wfau.roe.ac.uk/6dFGS/}.

The optical spectra of 2dFGRS galaxies were taken with the Two-degree Field (2dF) multi-object spectrograph 
on the Anglo-Australian Telescope.
This spectrograph has 140 $\mu$m diameter fibres corresponding to 2\arcsec.16 at the plate centre
and 1\arcsec.99 at the edge and covers 3600--8000 \AA\ with a spectral resolution of $\sim$ 9 \AA.
The 6dFGS galaxies were observed with the Six-degree Field (6dF) multi-object spectrograph 
having 6\arcsec.7 diameter fibres on the United Kingdom Schmidt Telescope.
The original grating in the 6dFGS spectrograph spans 4000--8400 \AA and gives a resolution of 5--12 \AA.
The improved grating, used after 2002 October, spans 3900--7500 \AA\ with a resolution of 4.9--6.6 \AA.
It is noted that the 2dFGRS and 6dFGS spectra are not properly flux-calibrated, but the flux ratio of adjacent lines is still useful
(to be discussed in Section 4).

In addition, we conducted optical spectroscopy of 15 ULIRGs in the survey sample plus one ULIRG, {\it IRAS} 09022$-$3615, in Sanders et al. (2003)
at the Cerro Tololo Inter-American Observatory (CTIO). The total number of our ULIRG sample with spectra was increased to 199.
The additional ULIRGs were observed on 2008 February 20--21 with the Ritchey-Chretien spectrograph
and the Loral 3K CCD  at the CTIO 4-m telescope (0\arcsec.5 pixel$^{-1}$).
A slit with a width of 1\arcsec.5 (a slightly larger than the seeing size) was adopted,
and was positioned in the east-west direction (P.A. = 90 \degr).
We used 316 lines mm$^{-1}$ grating to cover the spectral range 4500--10500 \AA\
with a resolution of 5.6 \AA\ ($\sim$ 2.8 pixels).
Three exposures were taken for each object and the integration times ranged from 360 to 3000 seconds depending on its brightness.

Data reduction was performed using the {\small IRAF} package.
This involved bias subtraction, flat fielding, sky subtraction, wavelength and flux calibration.
Galaxy spectra were extracted using an aperture width corresponding to a constant linear scale of 5 kpc
at the redshift of each galaxy.
A He-Ne-Ar lamp and standard star Hiltner 600 (Massey et al. 1988) taken nightly were used for wavelength and flux calibration, respectively.
In Fig. 1, we display 13 spectra with median signal-to-noise ratio (S/N) per pixel for the continuum greater than 3.
Typical emission lines (\Hb, \OIII$\lambda \lambda4959,5007$, \Ha+\NII$\lambda \lambda6548,6584$, and \SII$\lambda\lambda6717,6731$) 
are clearly seen in most spectra.

Among the sample of 199 ULIRGs considered in this study, we analyse the spectra of 115 ULIRGs with continuum S/N $>$ 3.
In the case that there are more than one spectrum for the same object,
the spectrum with a higher S/N of \Ha\ flux is chosen for the final classification.
If \Ha\ is not available from any spectra, the spectrum with a higher S/N of \Hb\ flux has priority.

\begin{figure*}
\begin{center}
\includegraphics [width=180mm] {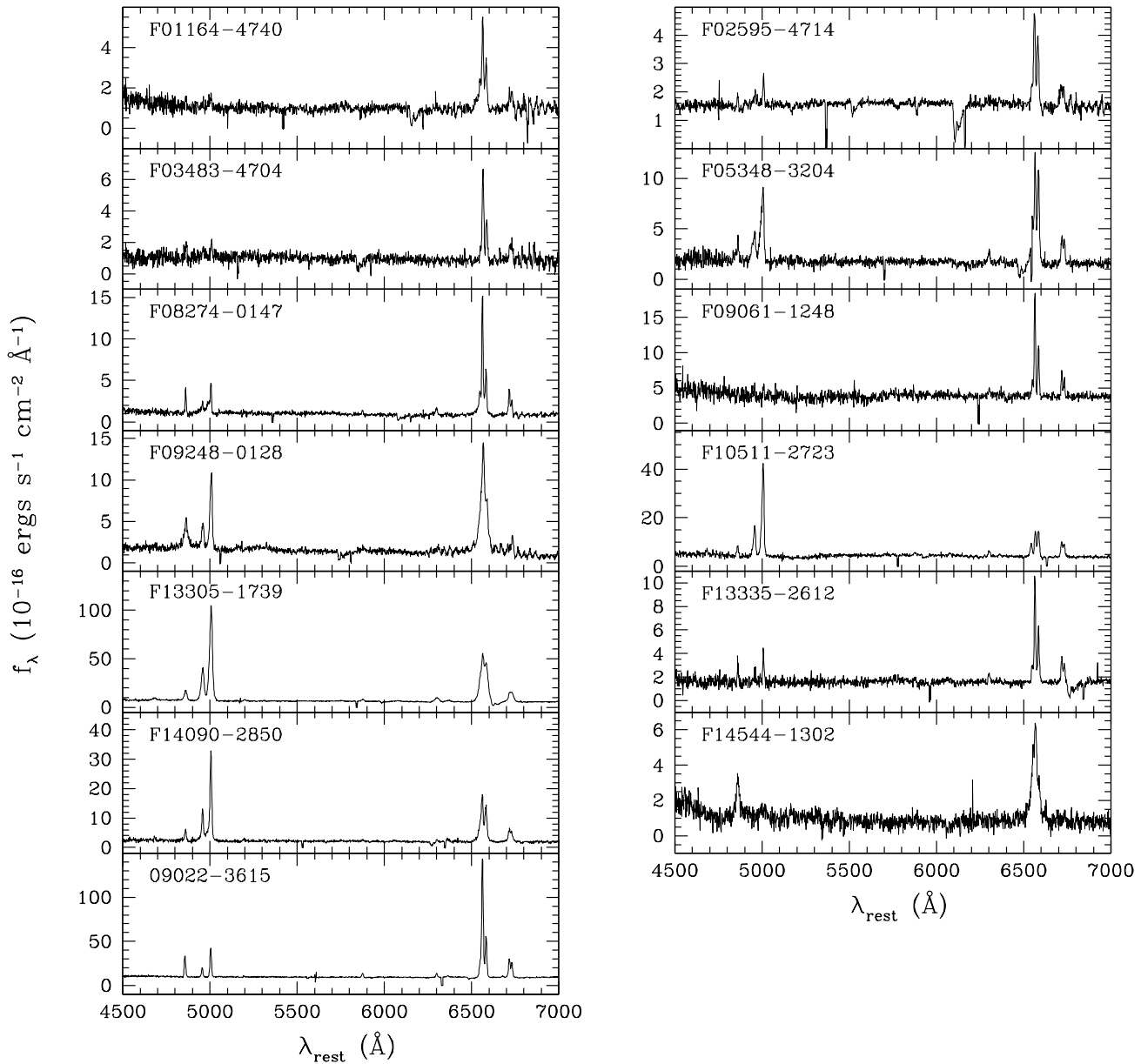}
\end{center}
\begin{flushleft}
\caption{Optical spectra of the ULIRGs observed at CTIO. These spectra are shifted to the rest-frame.
~~~~~~~~~~~~~~~~~~~~~~~~~~~~~~~~~~~~~}
\end{flushleft}
\end{figure*}

\begin{figure*}
\begin{center}
\includegraphics [width=140mm] {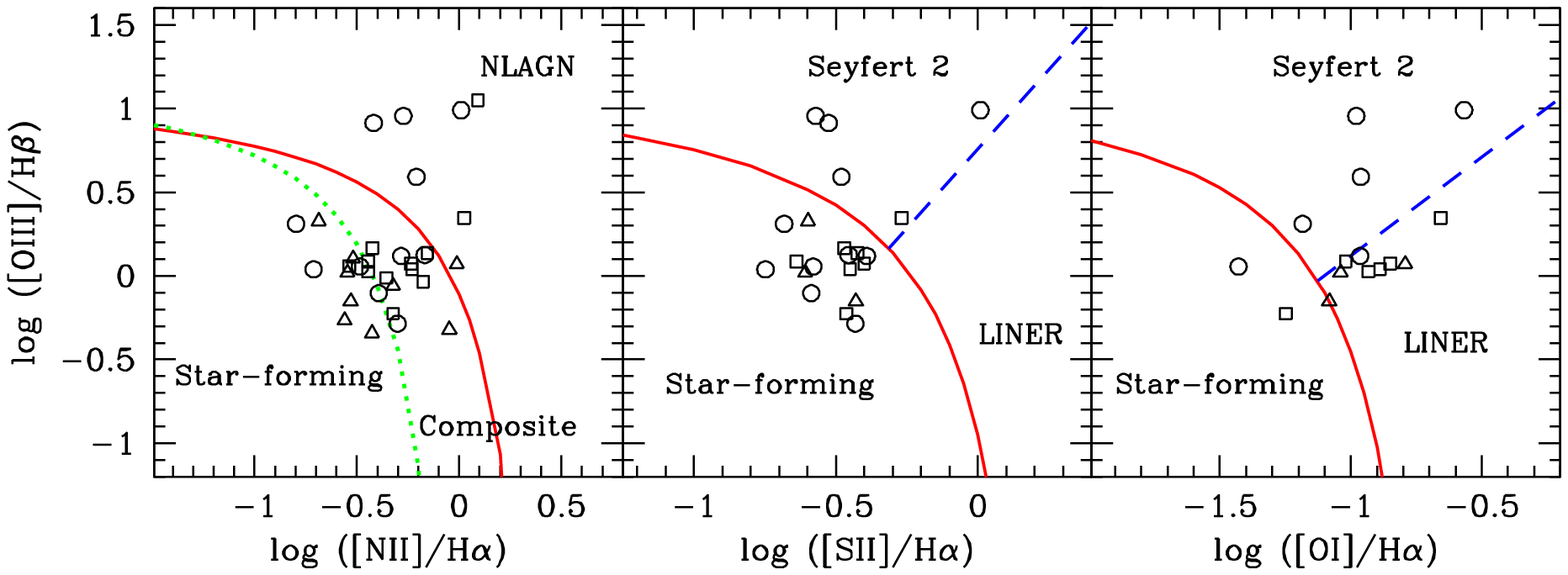}
\end{center}
\caption{The diagnostic diagrams for ULIRGs in the Sample A.
CTIO, 2dFGRS, and 6dFGS samples are represented by circle, square, and triangle, respectively.
The solid, dotted, and dashed lines indicate the extreme starburst (Kewley et al. 2001), pure star formation (Kauffmann et al. 2003),
and Seyfert-LINER (Kewley et al. 2006) lines, respectively.
NLAGN represents narrow-line AGN.}
\end{figure*}

\section{Analysis}

\subsection{Emission-line measurements}

After transforming the spectrum to the rest-frame and subtracting the local continuum defined by a linear fit around emission-lines,
we measure the line widths and fluxes via Gaussian profile fit using the {\small MPFIT/IDL} package
based on the Levenberg-Marquardt method (Markwardt 2009).
A single Gaussian is used to obtain the line width of \OIII$\lambda5007$.
For the line fluxes, we fit the \OI$\lambda6300$ line, \SII$\lambda\lambda6717,6731$ doublet, \Ha+\NII$\lambda \lambda6548,6584$
and \Hb+\OIII$\lambda \lambda4959,5007$ line complexes with one Gaussian, two Gaussians, four Gaussians, and six Gaussians, respectively.
The widths of the \NII\ doublet are kept the same and the height ratio of \NII$\lambda6548$ to \NII$\lambda6584$ is fixed to 1/3,
as required by the energy level structure of the \NII\ ion (Osterbrock \& Ferland 2006).
The \OIII\ doublet is fitted in the same way but each of the \OIII\ lines is modelled with two Gaussians
because \OIII\ line often has a blue, asymmetric wing that is perhaps from outflows of gas with opaque clouds (e.g., Heckman et al 1981; Greene \& Ho 2005).
Some AGN host galaxies show both narrow and broad Balmer lines (e.g., Osterbrock \& Mathews 1986; Hao et al. 2005)
and the narrow Balmer lines can have extended bases even in star-forming galaxies,
which are probably due to Wolf-Rayet stars (e.g., Osterbrook \& Cohen 1982; Brinchmann et al. 2008).
To take this into account, we fit each Balmer line with two Gaussians.
If the \Ha+\NII\ line complex cannot be decomposed into individual lines, the line fluxes are not kept.
Meanwhile, the decomposition of \SII\ doublet is not important because we are interested only in the total flux of the lines.
Uncertainties in the line measurements, comes from {\small MPFIT} routine, are typically 10--30 per cent.

\subsection{Corrections}

We correct the line fluxes for Galactic extinction using the foreground reddening maps provided by Schlegel et al. (1998)
and the extinction law of Cardelli et al. (1989).

Balmer emission lines can be strongly affected by stellar absorption features.
We remove the stellar absorption effects following the method discussed in Hopkins et al. (2003):
$S=F(EW+EW_c)/EW$, where $S$ is the stellar absorption-corrected line flux, $F$ is the measured line flux after foreground reddening correction,
$EW$ is the equivalent width of the line, and $EW_c$ is a correction factor.
We adopted $EW_c$ values (2.6 and 3.2 \AA\ for \Ha\ and \Hb, respectively) of Sc type galaxies (Miller \& Owen 2002).

If both \Ha\ and \Hb\ are measured, we can correct the line fluxes for internal extinction using the Balmer decrement
and the extinction curve with an assumption of an intrinsic \Ha/\Hb\ line ratio of 2.85 for star-forming galaxies and 3.1 for AGN galaxies
%(the Balmer decrement for case B recombination at T = 10$^{4}$ K and N$_{e} \sim $10$^{2}$--10$^{4}$ cm$^{-3}$; Osterbrock 1989).
(Osterbrock \& Ferland 2006).
We do not apply this correction when the observed ratio is smaller than the theoretical value.

An observed line width is the convolution of intrinsic line width and instrumental response.
Since the instrumental resolution of each survey is significantly different, the observed line width should be corrected 
for a fair comparison.
To correct the full width at half maximum (FWHM) of \OIII$\lambda5007$ line using the quadrature method,
we adopt 250 km s$^{-1}$ as a finite resolution of CTIO spectra.
Then we determine that the resolutions are roughly 450, 600, and 400 km s$^{-1}$ for the 2dFGRS, original 6dFGS, and improved 6dFGS grating,
respectively, by considering that the \OIII\ lines from different spectra have the same intrinsic width.
Note that corrected line widths below 100 km s$^{-1}$ should be treated with caution due to their large uncertainties ($>$ 200 km s$^{-1}$).

\subsection{Spectral classification}

\begin{figure}
\begin{center}
\includegraphics [width=80mm] {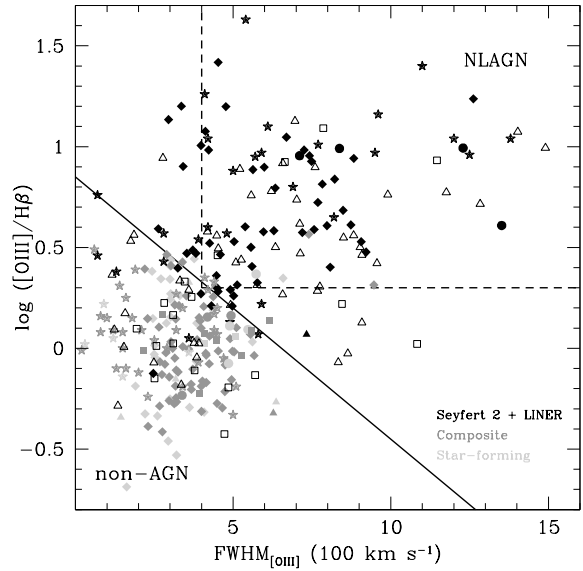}
\end{center}
\caption{log~(\OIII/\Hb) vs. FWHM$_{\scriptsize \OIII}$ diagram for the Sample B ULIRGs, which are denoted by open symbols.
The ULIRGs already classified in the diagnostic diagrams
are overplotted by filled symbols (black: Seyfert 2 plus LINER; dark-grey: composite; light-grey: star-forming).
Circles, squares, and triangles represent CTIO, 2dFGRS, and 6dFGS ULIRGs, respectively.
Stars and diamonds represent ULIRGs from the 1 Jy sample (Veilleux et al. 1999; Yuan et al. 2010)
and the SDSS sample (Hou et al. 2009), respectively.
The dashed and solid lines show the criteria of Zakamska et al. (2003) and our new boundary
separating narrow-line AGNs (NLAGNs) and non-AGNs.
In this diagram, \OIII/\Hb\ line ratios are corrected only for Galactic extinction and stellar absorption.}
\end{figure}

\begin{table*}
\begin{flushleft}
\textbf{Table 1.} Basic information and spectral types of 115 southern ULIRGs.\\
\end{flushleft}
\begin{center}
\begin{tabular}{rcccrrrclccl}
\hline
{\it IRAS} name & RA (J2000) & Dec (J2000) & z & \multicolumn{1}{c}{$f_{25}$} & \multicolumn{1}{c}{$f_{60}$} &
\multicolumn{1}{c}{$f_{100}$} & \multicolumn{1}{c}{$\log~\dfrac{L_{\mbox{\scriptsize IR}}}{L_{\odot}}$} &
\multicolumn{1}{c}{sample} & class & sub. &  \multicolumn{1}{c}{known}\\
\hline
 F00050$-$3259 & 00 07 34.6 & $-$32 43 03 & 0.285 & $<$0.144 &  0.222 & $<$0.758 & 12.12 &  2dFGRS &  X  &                &                  \\
 F00091$-$3905 & 00 11 42.3 & $-$38 49 15 & 0.253 & $<$0.125 &  0.316 & $<$0.756 & 12.14 &  6dFGS  &  N  &                &                  \\
 F00184$-$3331 & 00 20 57.7 & $-$33 14 28 & 0.238 & $<$0.120 &  0.334 &    0.613 & 12.10 &  2dFGRS &  X  &                &                  \\
 F00318$-$3137 & 00 34 16.0 & $-$31 21 04 & 0.284 & $<$0.167 &  0.257 &    0.563 & 12.18 &  6dFGS  &  X  &                &                  \\
 F00335$-$2732 & 00 35 59.2 & $-$27 15 42 & 0.068 &    0.632 &  4.294 &    3.207 & 12.01 &  2dFGRS &  X  &           Co   &            Co(5) \\
 F00456$-$2904 & 00 48 03.5 & $-$28 48 38 & 0.110 &    0.141 &  2.598 &    3.377 & 12.23 &  2dFGRS &  X  &           Co   &            Co(5) \\
 F00482$-$2721 & 00 50 40.0 & $-$27 04 42 & 0.129 & $<$0.182 &  1.134 &    1.839 & 12.03 &  2dFGRS &  X  &           Co   &            Co(5) \\
 F00569$-$3108 & 00 59 22.3 & $-$30 52 27 & 0.344 & $<$0.055 &  0.239 & $<$0.463 & 12.34 &  6dFGS  &  N  &                &                  \\
 F01004$-$2237 & 01 02 51.2 & $-$22 21 50 & 0.117 &    0.660 &  2.287 &    1.790 & 12.24 &  6dFGS  &  N  &                &      Co(5)$^{a}$ \\
 F01009$-$3241 & 01 03 19.8 & $-$32 25 37 & 0.256 & $<$0.123 &  0.291 & $<$0.750 & 12.12 &  2dFGRS &  N  &                &                  \\
 F01160$-$2551 & 01 18 24.2 & $-$25 36 19 & 0.237 & $<$0.138 &  0.452 &    1.153 & 12.23 &  2dFGRS &  X  &                &                  \\
 F01164$-$4740 & 01 18 36.5 & $-$47 25 04 & 0.235 & $<$0.219 &  0.487 &    0.674 & 12.25 &  CTIO   &  X  &           SF   &                  \\
 F01212$-$5025 & 01 23 20.1 & $-$50 09 29 & 0.201 & $<$0.077 &  0.415 &    0.717 & 12.02 &  6dFGS  &  N  &                &                  \\
 F01234$-$0447 & 01 25 56.0 & $-$04 31 57 & 0.156 & $<$0.162 &  0.711 &    1.108 & 12.01 &  6dFGS  &  X  &                &                  \\
 F01358$-$3300 & 01 38 05.3 & $-$32 45 29 & 0.197 & $<$0.138 &  0.835 &    1.079 & 12.31 &  2dFGRS &  X  &                &            SF(1) \\
 F01379$-$3203 & 01 40 15.3 & $-$31 48 18 & 0.202 & $<$0.146 &  0.686 &    0.929 & 12.25 &  6dFGS  &  N  &                &                  \\
 F01497$-$2906 & 01 52 04.0 & $-$28 51 18 & 0.183 & $<$0.052 &  0.588 &    1.278 & 12.08 &  2dFGRS &  X  &           Co   &                  \\
 F01569$-$2939 & 01 59 13.1 & $-$29 24 37 & 0.140 &    0.143 &  1.734 &    1.514 & 12.29 &  2dFGRS &  N  &                &      Co(5)$^{a}$ \\
 F02038$-$0816 & 02 06 18.7 & $-$08 02 32 & 0.220 & $<$0.326 &  0.351 &    0.525 & 12.04 &  6dFGS  &  N  &                &                  \\
 F02068$-$2000 & 02 09 09.0 & $-$19 46 30 & 0.253 & $<$0.190 &  0.278 & $<$0.647 & 12.09 &  6dFGS  &  N  &                &                  \\
 F02130$-$1948 & 02 15 23.5 & $-$19 34 18 & 0.191 & $<$0.196 &  0.553 &    0.565 & 12.10 &  6dFGS  &  N  &                &                  \\
 F02356$-$4628 & 02 37 29.3 & $-$46 15 48 & 0.206 & $<$0.090 &  1.034 &    1.738 & 12.45 &  2dFGRS &  X  &                &                  \\
 F02361$-$3233 & 02 38 15.2 & $-$32 20 34 & 0.198 & $<$0.073 &  0.741 &    1.821 & 12.26 &  2dFGRS &  X  &           SF   &                  \\
 F02364$-$4751 & 02 38 13.6 & $-$47 38 06 & 0.097 &    0.175 &  2.794 &    4.953 & 12.15 &  6dFGS  &  N  &                &                  \\
 F02384$-$1744 & 02 40 46.4 & $-$17 31 52 & 0.308 & $<$0.072 &  0.269 & $<$1.344 & 12.28 &  6dFGS  &  X  &                &                  \\
 F02437$-$1145 & 02 46 07.8 & $-$11 32 42 & 0.270 &    0.148 &  0.208 & $<$0.710 & 12.03 &  6dFGS  &  B  &           S1   &                  \\
 F02595$-$4714 & 03 01 18.4 & $-$47 02 17 & 0.245 & $<$0.063 &  0.268 &    0.484 & 12.04 &  CTIO   &  X  &           Co   &                  \\
 F03000$-$2719 & 03 02 10.6 & $-$27 07 29 & 0.221 & $<$0.113 &  0.918 &    2.039 & 12.47 &  2dFGRS &  N  &                &                  \\
 F03130$-$3119 & 03 15 04.6 & $-$31 08 02 & 0.258 & $<$0.103 &  0.252 &    0.438 & 12.06 &  2dFGRS &  X  &                &                  \\
 F03259$-$3105 & 03 27 59.3 & $-$30 54 50 & 0.261 & $<$0.065 &  0.249 &    0.559 & 12.07 &  2dFGRS &  X  &                &                  \\
 F03483$-$4704 & 03 49 54.4 & $-$46 55 10 & 0.301 & $<$0.098 &  0.156 & $<$0.542 & 12.02 &  CTIO   &  X  &           SF   &                  \\
 F03485$-$1827 & 03 50 45.7 & $-$18 18 27 & 0.174 &    0.164 &  0.580 & $<$0.822 & 12.03 &  6dFGS  &  N  &                &                  \\
 F03569$-$2535 & 03 59 00.9 & $-$25 26 44 & 0.220 &    0.071 &  0.613 &    0.598 & 12.28 &  6dFGS  &  X  &                &                  \\
 F04056$-$5722 & 04 06 41.9 & $-$57 14 38 & 0.267 &    0.079 &  0.259 &    0.453 & 12.11 &  6dFGS  &  X  &                &                  \\
 F04279$-$3035 & 04 29 55.7 & $-$30 28 54 & 0.232 & $<$0.114 &  0.311 & $<$0.839 & 12.04 &  6dFGS  &  X  &                &                  \\
 F04489$-$1026 & 04 51 19.5 & $-$10 21 22 & 0.231 & $<$0.090 &  0.357 & $<$1.767 & 12.10 &  6dFGS  &  N  &                &                  \\
 F04505$-$2958 & 04 52 30.7 & $-$29 53 34 & 0.285 &    0.189 &  0.650 &    0.765 & 12.58 &  6dFGS  &  B  &           S1   &            S1(4) \\
 F05020$-$2941 & 05 04 00.8 & $-$29 36 57 & 0.154 &    0.102 &  1.932 &    2.059 & 12.43 &  6dFGS  &  X  &                &            Co(5) \\
 F05156$-$3024 & 05 17 31.9 & $-$30 21 14 & 0.171 &    0.103 &  1.162 &    1.402 & 12.31 &  6dFGS  &  N  &                &            S2(5) \\
 F05348$-$3204 & 05 36 44.9 & $-$32 02 20 & 0.174 &    0.224 &  0.623 &    0.712 & 12.06 &  CTIO   &  N  &           S2   &                  \\
 F06158$-$4017 & 06 17 29.8 & $-$40 18 57 & 0.221 &    0.083 &  0.311 & $<$0.707 & 12.00 &  6dFGS  &  N  &                &                  \\
 F08274$-$0147 & 08 30 00.6 & $-$01 57 05 & 0.250 &    0.149 &  0.594 &    0.622 & 12.40 &  CTIO   &  X  &           SF   &                  \\
 F08411$-$2501 & 08 43 18.8 & $-$25 11 58 & 0.134 &    0.262 &  1.655 &    1.572 & 12.23 &  6dFGS  &  N  &                &                  \\
 F09061$-$1248 & 09 08 35.1 & $-$13 01 01 & 0.073 &    0.191 &  3.634 &    5.316 & 12.01 &  CTIO   &  X  &           Co   &            SF(2) \\
 F09090$-$1349 & 09 11 27.9 & $-$14 01 40 & 0.171 & $<$0.091 &  0.845 &    1.553 & 12.17 &  6dFGS  &  X  &                &                  \\
 F09248$-$0128 & 09 27 23.9 & $-$01 41 19 & 0.324 & $<$0.102 &  0.221 & $<$0.667 & 12.25 &  CTIO   &  B  &           S1   &                  \\
 F09521$-$0400 & 09 54 37.2 & $-$04 15 12 & 0.237 & $<$0.108 &  0.387 &    0.600 & 12.16 &  2dFGRS &  N  &                &                  \\
 F10077$+$0034 & 10 10 16.6 &$~~~$00 19 31& 0.182 & $<$0.127 &  0.513 &    0.834 & 12.02 &  2dFGRS &  X  &           SF   &                  \\
 F10298$-$2300 & 10 32 11.7 & $-$23 15 41 & 0.285 & $<$0.153 &  0.377 & $<$1.583 & 12.35 &  6dFGS  &  N  &                &                  \\
 F10479$-$2808 & 10 50 18.9 & $-$28 24 01 & 0.191 &    0.327 &  1.002 &    1.151 & 12.36 &  6dFGS  &  B  &           S1   &                  \\
 F10511$-$2723 & 10 53 33.3 & $-$27 39 05 & 0.159 &    0.319 &  0.896 & $<$1.351 & 12.13 &  CTIO   &  N  &           S2   &                  \\
 F10533$-$3534 & 10 55 40.2 & $-$35 50 06 & 0.190 & $<$0.096 &  0.556 & $<$1.153 & 12.09 &  6dFGS  &  X  &           Co   &                  \\
 F10549$-$3702 & 10 57 18.5 & $-$37 18 25 & 0.216 & $<$0.077 &  0.348 & $<$1.265 & 12.02 &  6dFGS  &  N  &                &                  \\
 F11053$-$2413 & 11 07 47.0 & $-$24 29 25 & 0.225 & $<$0.092 &  0.314 & $<$2.103 & 12.02 &  6dFGS  &  N  &                &                  \\
 F11093$-$3353 & 11 11 45.4 & $-$34 09 35 & 0.231 &    0.102 &  0.318 &    0.894 & 12.05 &  6dFGS  &  X  &           SF   &                  \\
 F11095$-$0238 & 11 12 02.5 & $-$02 54 18 & 0.106 &    0.418 &  3.249 &    2.531 & 12.30 &  2dFGRS &  X  &           Co   &            Co(5) \\
 F11204$-$2154 & 11 22 58.0 & $-$22 11 02 & 0.248 & $<$0.114 &  0.264 & $<$1.263 & 12.04 &  6dFGS  &  X  &                &                  \\
 F11300$-$0522 & 11 32 41.4 & $-$05 39 41 & 0.230 & $<$0.164 &  0.668 &    1.610 & 12.37 &  2dFGRS &  X  &                &                  \\
 F11451$-$2128 & 11 47 39.0 & $-$21 45 06 & 0.219 & $<$0.182 &  0.375 & $<$0.859 & 12.07 &  6dFGS  &  B  &           S1   &                  \\
 F12131$-$2809 & 12 15 42.4 & $-$28 26 19 & 0.360 & $<$0.102 &  0.325 & $<$1.811 & 12.52 &  6dFGS  &  N  &                &                  \\
 F12432$-$3138 & 12 45 57.2 & $-$31 54 42 & 0.423 & $<$0.154 &  0.256 & $<$0.718 & 12.61 &  6dFGS  &  N  &                &                  \\
 F12452$-$2032 & 12 47 53.6 & $-$20 48 24 & 0.209 & $<$0.203 &  0.383 & $<$0.988 & 12.03 &  6dFGS  &  X  &                &                  \\
 F13269$-$2251 & 13 29 40.8 & $-$23 07 10 & 0.290 & $<$0.114 &  0.382 & $<$1.410 & 12.37 &  6dFGS  &  X  &                &                  \\
\hline
\end{tabular}
\end{center}
\end{table*}
\begin{table*}
\begin{flushleft}
\textbf{Table 1.} -- Continued. \\
\end{flushleft}
\begin{center}
\begin{tabular}{rcccrrrclccl}
\hline
{\it IRAS} name & RA (J2000) & Dec (J2000) & z & \multicolumn{1}{c}{$f_{25}$} & \multicolumn{1}{c}{$f_{60}$} &
\multicolumn{1}{c}{$f_{100}$} &  \multicolumn{1}{c}{$\log~\dfrac{L_{\mbox{\scriptsize IR}}}{L_{\odot}}$} &
\multicolumn{1}{c}{sample} & class & sub. &  \multicolumn{1}{c}{known}\\
\hline
 F13270$-$0331 & 13 29 40.7 & $-$03 46 59 & 0.221 & $<$0.306 &  0.954 &    0.797 & 12.48 &  2dFGRS &  N  &                &                  \\
 F13305$-$1739 & 13 33 15.2 & $-$17 55 00 & 0.148 &    0.392 &  1.164 &    1.044 & 12.17 &  CTIO   &  N  &           S2   &            S2(5) \\
 F13306$-$1644 & 13 33 21.5 & $-$17 00 22 & 0.231 & $<$0.208 &  0.298 & $<$0.752 & 12.02 &  6dFGS  &  X  &                &                  \\
 F13335$-$2612 & 13 36 22.1 & $-$26 27 30 & 0.125 & $<$0.139 &  1.402 &    2.101 & 12.09 &  CTIO   &  X  &           Co   &            Co(5) \\
 F13379$-$0256 & 13 40 33.4 & $-$03 11 42 & 0.218 & $<$0.235 &  0.728 &    1.032 & 12.35 &  2dFGRS &  X  &                &                  \\
 F13531$-$3422 & 13 56 06.6 & $-$34 37 02 & 0.220 &    0.135 &  0.380 & $<$0.777 & 12.08 &  6dFGS  &  N  &                &                  \\
 F14021$-$3139 & 14 05 02.1 & $-$31 54 17 & 0.202 & $<$0.178 &  0.568 & $<$1.124 & 12.17 &  6dFGS  &  N  &                &                  \\
 F14090$-$2850 & 14 11 59.2 & $-$29 05 01 & 0.212 &    0.197 &  0.507 & $<$0.807 & 12.17 &  CTIO   &  N  &           S2   &                  \\
 F14121$-$0126 & 14 14 45.7 & $-$01 40 53 & 0.150 & $<$0.239 &  1.394 &    2.073 & 12.26 &  2dFGRS &  X  &           Co   &      S2(5)$^{a}$ \\
 F14207$-$2002 & 14 23 31.5 & $-$20 15 47 & 0.173 & $<$0.208 &  0.850 &    1.082 & 12.18 &  6dFGS  &  N  &                &            S2(1) \\
 F14248$-$3644 & 14 27 51.7 & $-$36 58 03 & 0.208 & $<$0.122 &  0.450 & $<$1.499 & 12.10 &  6dFGS  &  X  &                &                  \\
 F14254$-$2655 & 14 28 19.9 & $-$27 08 49 & 0.253 & $<$0.206 &  0.534 &    0.809 & 12.37 &  6dFGS  &  N  &                &            S2(1) \\
 F14348$-$1447 & 14 37 37.2 & $-$15 00 20 & 0.082 &    0.495 &  6.870 &    7.068 & 12.39 &  6dFGS  &  X  &           SF   &      Co(5)$^{a}$ \\
 F14544$-$1302 & 14 57 09.5 & $-$13 14 55 & 0.254 & $<$0.265 &  0.309 & $<$0.761 & 12.13 &  CTIO   &  B  &           S1   &                  \\
 F15130$-$1958 & 15 15 55.5 & $-$20 09 17 & 0.108 &    0.388 &  1.916 &    2.299 & 12.09 &  6dFGS  &  N  &                &            S2(3) \\
 F16090$-$0139 & 16 11 40.9 & $-$01 47 06 & 0.134 &    0.264 &  3.609 &    4.874 & 12.57 &  6dFGS  &  X  &           Co   &      Co(3)$^{b}$ \\
 F16159$-$0402 & 16 18 36.3 & $-$04 09 42 & 0.211 &    0.299 &  0.979 & $<$1.768 & 12.45 &  6dFGS  &  N  &                &                  \\
 F19466$-$3649 & 19 49 55.5 & $-$36 42 06 & 0.093 &    0.316 &  2.425 &    3.378 & 12.05 &  6dFGS  &  X  &           SF   &                  \\
 F19548$-$6237 & 19 59 18.9 & $-$62 29 18 & 0.351 & $<$0.081 &  0.365 & $<$0.823 & 12.55 &  6dFGS  &  X  &                &                  \\
 F20023$-$5253 & 20 06 08.1 & $-$52 44 48 & 0.238 & $<$0.089 &  0.312 & $<$1.454 & 12.07 &  6dFGS  &  X  &                &                  \\
 F20066$-$1630 & 20 09 27.7 & $-$16 22 06 & 0.163 & $<$0.146 &  0.639 & $<$1.285 & 12.00 &  6dFGS  &  X  &                &                  \\
 F20181$-$2244 & 20 21 03.9 & $-$22 35 22 & 0.184 & $<$0.199 &  0.568 &    0.885 & 12.07 &  6dFGS  &  N  &                &                  \\
 F20248$-$3204 & 20 27 59.1 & $-$31 54 54 & 0.203 & $<$0.119 &  0.440 & $<$2.673 & 12.06 &  6dFGS  &  X  &                &                  \\
 F20270$-$4237 & 20 30 24.9 & $-$42 27 24 & 0.242 & $<$0.102 &  0.287 & $<$0.646 & 12.05 &  6dFGS  &  N  &                &                  \\
 F20273$-$6558 & 20 31 50.8 & $-$65 48 22 & 0.349 &    0.155 &  0.430 & $<$1.151 & 12.61 &  6dFGS  &  N  &                &                  \\
 F20542$-$1832 & 20 57 03.6 & $-$18 20 43 & 0.298 & $<$0.126 &  0.322 &    0.934 & 12.33 &  6dFGS  &  N  &                &                  \\
 F20551$-$4250 & 20 58 27.3 & $-$42 38 57 & 0.042 &    1.906 & 12.780 &    9.948 & 12.06 &  6dFGS  &  X  &           SF   &      Co(5)$^{a}$ \\
 F21016$-$1900 & 21 04 29.8 & $-$18 48 17 & 0.230 & $<$0.230 &  0.305 &    1.025 & 12.03 &  6dFGS  &  N  &                &                  \\
 F21356$-$1015 & 21 38 20.2 & $-$10 01 57 & 0.206 &    0.159 &  0.460 &    0.550 & 12.09 &  6dFGS  &  N  &                &                  \\
 F21367$-$2405 & 21 39 36.6 & $-$23 51 51 & 0.234 & $<$0.102 &  0.383 &    0.565 & 12.15 &  6dFGS  &  X  &           SF   &                  \\
 F21435$-$3648 & 21 46 31.9 & $-$36 34 54 & 0.160 & $<$0.149 &  0.665 &    0.999 & 12.01 &  6dFGS  &  N  &                &                  \\
 F21488$-$2819 & 21 51 41.4 & $-$28 05 14 & 0.234 & $<$0.135 &  0.301 & $<$0.677 & 12.04 &  2dFGRS &  X  &                &                  \\
 F21542$-$4050 & 21 57 21.3 & $-$40 36 03 & 0.301 & $<$0.093 &  0.253 & $<$0.531 & 12.23 &  6dFGS  &  B  &           S1   &                  \\
 F21555$-$4235 & 21 58 37.5 & $-$42 21 33 & 0.181 & $<$0.132 &  0.833 &    0.987 & 12.22 &  6dFGS  &  X  &                &                  \\
 F22058$-$3501 & 22 08 48.8 & $-$34 46 38 & 0.173 & $<$0.162 &  0.561 &    1.375 & 12.01 &  2dFGRS &  N  &                &                  \\
 F22206$-$2715 & 22 23 29.4 & $-$26 59 59 & 0.131 & $<$0.159 &  1.754 &    2.333 & 12.23 &  2dFGRS &  X  &           Co   &            Co(5) \\
 F22301$-$2822 & 22 32 56.7 & $-$28 07 17 & 0.244 & $<$0.123 &  0.312 & $<$0.925 & 12.10 &  2dFGRS &  N  &                &                  \\
 F22423$-$4707 & 22 45 20.2 & $-$46 52 03 & 0.200 &    0.150 &  0.451 &    0.693 & 12.05 &  6dFGS  &  B  &           S1   &                  \\
 F22521$-$3929 & 22 54 56.5 & $-$39 13 14 & 0.261 &    0.135 &  0.283 & $<$0.545 & 12.13 &  6dFGS  &  N  &                &                  \\
 F22546$-$2637 & 22 57 23.8 & $-$26 21 23 & 0.163 & $<$0.166 &  0.752 &    1.362 & 12.08 &  2dFGRS &  X  &                &            SF(1) \\
 F22560$-$3501 & 22 58 46.7 & $-$34 45 44 & 0.171 &    0.141 &  0.593 &    0.795 & 12.02 &  2dFGRS &  X  &           Co   &                  \\
 F23046$-$3454 & 23 07 21.3 & $-$34 38 41 & 0.208 & $<$0.093 &  0.937 &    1.313 & 12.41 &  2dFGRS &  X  &                &                  \\
 F23128$-$5919 & 23 15 46.5 & $-$59 03 14 & 0.044 &    1.590 & 10.800 &   10.990 & 12.03 &  6dFGS  &  X  &           SF   &                  \\
 F23142$-$0611 & 23 16 49.3 & $-$05 55 13 & 0.346 & $<$0.158 &  0.263 & $<$0.405 & 12.39 &  6dFGS  &  N  &                &                  \\
 F23185$-$0328 & 23 21 05.9 & $-$03 12 03 & 0.246 & $<$0.228 &  0.308 & $<$0.385 & 12.10 &  6dFGS  &  N  &                &                  \\
 F23206$-$1222 & 23 23 14.4 & $-$12 06 29 & 0.249 & $<$0.255 &  0.246 & $<$0.689 & 12.01 &  2dFGRS &  N  &                &                  \\
 F23242$-$0357 & 23 26 49.1 & $-$03 41 18 & 0.189 & $<$0.275 &  0.454 &    0.566 & 12.00 &  6dFGS  &  X  &                &                  \\
 F23253$-$5415 & 23 28 06.0 & $-$53 58 26 & 0.129 &    0.214 &  2.296 &    3.493 & 12.34 &  6dFGS  &  N  &           LI   &                  \\
 F23516$-$2420 & 23 54 13.0 & $-$24 04 05 & 0.154 & $<$0.283 &  0.834 &    1.240 & 12.06 &  6dFGS  &  X  &                &                  \\
 F23529$-$2119 & 23 55 33.8 & $-$21 02 49 & 0.428 & $<$0.156 &  0.327 &    0.627 & 12.73 &  6dFGS  &  N  &                &                  \\
 F23559$-$3009 & 23 58 31.0 & $-$29 52 18 & 0.342 & $<$0.152 &  0.237 & $<$0.463 & 12.33 &  2dFGRS &  X  &                &                  \\
  09022$-$3615 & 09 04 12.8 & $-$36 27 02 & 0.059 &    1.154 & 11.470 &   11.080 & 12.26 &  CTIO   &  X  &           SF   &                  \\
\hline
\end{tabular}
\begin{flushleft}
Column descriptions:
(1) Object name in the {\it IRAS} catalogue.
(2-3) Right ascension and declination in units of $^{h~m~s}$ and \degr\ \arcmin\ \arcsec, respectively.
(4) Redshift.
(5-7) The {\it IRAS} flux density at 25, 60, and 100 $\mu$m [Jy].
(8) Infrared luminosity.
(1-8) Basic information taken from Hwang et al. (2007) except 09022$-$3615, for which information was taken from Sanders et al. (2003).
(9) Adopted spectrum.
(10) Spectral class in this study (B$=$broad-line AGN, N$=$narrow-line AGN, X$=$non-AGN).
(11) Subclass in this study (S1=Seyfert 1, S2=Seyfert 2, LI=LINER, Co=composite, SF=star-forming galaxy).
(12) Subclass from previous studies. Numbers in parentheses are references
(1=Allen et al. 1991, 2=Duc et al. 1997, 3=Kim et al. 1998, 4=Low et al. 1988, 5=Yuan et al. 2010).\\
$^{a}$   The spectral types of these objects are different between this study and previous studies.\\
$^{b}$   This object was classified as a LINER by Kim et al. (1998) using the Veilleux \& Osterbrock (1987) scheme.
\end{flushleft}
\end{center}
\end{table*}

\begin{table*}
\begin{flushleft}
\textbf{Table 2.} Line information of the Sample A ULIRGs.\\
\end{flushleft}
\begin{center}
\begin{tabular}{rcrrr@{.}lr@{}lr@{}lr@{}lr@{}lr@{}l}
\hline
\multicolumn{1}{c}{{\it IRAS} name} & quality &
\multicolumn{1}{c}{$f_{\scriptsize \Ha}$}  &
\multicolumn{1}{r}{EW$_{\scriptsize \Ha}$}  &
\multicolumn{2}{c}{$\dfrac{\Ha}{\Hb}$} &
\multicolumn{2}{c}{FWHM$_{\scriptsize \OIII}$}    &
\multicolumn{2}{c}{~$\log \dfrac{\OIII}{\Hb}$} & \multicolumn{2}{c}{~$\log \dfrac{\NII}{\Ha}$} &
\multicolumn{2}{c}{~$\log \dfrac{\SII}{\Ha}$}  & \multicolumn{2}{c}{~$\log \dfrac{\OI}{\Ha}$}\\
\hline
F00335$-$2732 & 3333333 &   ...  &    50 &  5 & 81  &    559&$\pm$23  &   0.06 & $\pm$0.10  & $-$0.44 & $\pm$0.06  & $-$0.66 & $\pm$0.10  & $-$0.99 &$\pm$0.10  \\
F00456$-$2904 & 3333333 &   ...  &    78 &  4 & 75  &    394&$\pm$16  &$-$0.25 & $\pm$0.03  & $-$0.33 & $\pm$0.02  & $-$0.48 & $\pm$0.04  & $-$1.23 &$\pm$0.09  \\
F00482$-$2721 & 2323320 &   ...  &    40 &  5 & 44  &    445&$\pm$76  &   0.01 & $\pm$0.19  & $-$0.23 & $\pm$0.06  & $-$0.49 & $\pm$0.30  & $-$0.86 &$\pm$0.22  \\
F01164$-$4740 & 1103322 &    8.5 &    86 & 10 & 88  &    489&$\pm$168 &   0.04 & $\pm$0.45  & $-$0.71 & $\pm$0.16  & $-$0.75 & $\pm$0.25  &         &...        \\
F01497$-$2906 & 3203300 &   ...  &    47 &  4 & 09  &    526&$\pm$74  &$-$0.06 & $\pm$0.23  & $-$0.18 & $\pm$0.12  &         & ...        &         &...        \\
F01569$-$2939 & 3333333 &   ...  &    97 &  5 & 58  &    636&$\pm$8   &   0.32 & $\pm$0.03  &    0.02 & $\pm$0.02  & $-$0.28 & $\pm$0.06  & $-$0.63 &$\pm$0.03  \\
F02361$-$3233 & 3303300 &   ...  &    72 &  5 & 14  &    507&$\pm$21  &   0.03 & $\pm$0.10  & $-$0.54 & $\pm$0.10  &         & ...        &         &...        \\
F02595$-$4714 & 2303321 &    3.9 &    25 &  5 & 69  &    493&$\pm$45  &   0.12 & $\pm$0.18  & $-$0.17 & $\pm$0.06  & $-$0.46 & $\pm$0.29  &         &...        \\
F03483$-$4704 & 1203311 &    6.5 &    72 &  4 & 78  &    485&$\pm$64  &$-$0.10 & $\pm$0.35  & $-$0.39 & $\pm$0.10  & $-$0.59 & $\pm$0.35  &         &...        \\
F05348$-$3204 & 1323322 &   16.5 &    99 &  4 & 24  &   1351&$\pm$38  &   0.59 & $\pm$0.22  & $-$0.21 & $\pm$0.06  & $-$0.48 & $\pm$0.21  & $-$0.96 &$\pm$0.25  \\
F08274$-$0147 & 3323333 &   20.4 &   236 &  8 & 18  &    573&$\pm$88  &   0.31 & $\pm$0.08  & $-$0.80 & $\pm$0.07  & $-$0.68 & $\pm$0.11  & $-$1.18 &$\pm$0.20  \\
F09061$-$1248 & 1103332 &   13.4 &    36 &  7 & 72  &    338&$\pm$145 &$-$0.29 & $\pm$0.38  & $-$0.30 & $\pm$0.10  & $-$0.43 & $\pm$0.23  &         &...        \\
F10077$+$0034 & 2222300 &   ...  &    66 &  4 & 55  &    410&$\pm$56  &   0.03 & $\pm$0.25  & $-$0.44 & $\pm$0.21  &         & ...        & $-$0.93 &$\pm$0.30  \\
F10511$-$2723 & 2323332 &   12.5 &    25 &  2 & 26  &    837&$\pm$10  &   0.99 & $\pm$0.13  &    0.01 & $\pm$0.07  &    0.01 & $\pm$0.17  & $-$0.57 &$\pm$0.19  \\
F10533$-$3534 & 2103300 &   ...  &   201 &  1 & 71  &       &...      &$-$0.06 & $\pm$0.46  & $-$0.32 & $\pm$0.17  &         & ...        &         &...        \\
F11093$-$3353 & 2102200 &   ...  &    36 &  1 & 31  &    636&$\pm$157 &$-$0.27 & $\pm$0.40  & $-$0.56 & $\pm$0.31  &         & ...        &         &...        \\
F11095$-$0238 & 3232322 &   ...  &    54 &  5 & 39  &    409&$\pm$40  &   0.07 & $\pm$0.28  & $-$0.24 & $\pm$0.26  & $-$0.40 & $\pm$0.29  & $-$0.85 &$\pm$0.30  \\
F13305$-$1739 & 3333330 &  128.5 &   228 &  6 & 15  &   1229&$\pm$9   &   0.96 & $\pm$0.08  & $-$0.27 & $\pm$0.05  & $-$0.57 & $\pm$0.04  & $-$0.98 &$\pm$0.06  \\
F13335$-$2612 & 2323321 &    9.0 &    58 &  5 & 66  &    418&$\pm$52  &   0.12 & $\pm$0.20  & $-$0.28 & $\pm$0.20  & $-$0.39 & $\pm$0.38  & $-$0.96 &$\pm$0.37  \\
F14090$-$2850 & 3303332 &   33.3 &   161 &  6 & 76  &    710&$\pm$12  &   0.91 & $\pm$0.08  & $-$0.42 & $\pm$0.09  & $-$0.53 & $\pm$0.18  &         &...        \\
F14121$-$0126 & 2103332 &   ...  &    68 & 10 & 40  &    615&$\pm$51  &   0.07 & $\pm$0.41  & $-$0.16 & $\pm$0.13  & $-$0.46 & $\pm$0.18  &         &...        \\
F14348$-$1447 & 3333333 &   ...  &    99 &  4 & 68  &       &...      &$-$0.01 & $\pm$0.11  & $-$0.55 & $\pm$0.09  & $-$0.62 & $\pm$0.16  & $-$1.02 &$\pm$0.20  \\
F16090$-$0139 & 2103300 &   ...  &    27 &  2 & 13  &    627&$\pm$176 &$-$0.32 & $\pm$0.40  & $-$0.05 & $\pm$0.17  &         & ...        &         &...        \\
F19466$-$3649 & 2103300 &   ...  &    15 &  3 & 78  &    143&$\pm$120 &$-$0.37 & $\pm$0.43  & $-$0.42 & $\pm$0.10  &         & ...        &         &...        \\
F20551$-$4250 & 3333333 &   ...  &    60 &  3 & 92  &    318&$\pm$13  &$-$0.17 & $\pm$0.03  & $-$0.53 & $\pm$0.04  & $-$0.44 & $\pm$0.05  & $-$1.07 &$\pm$0.06  \\
F21367$-$2405 & 3303300 &   ...  &    91 &  1 & 23  &       &...      &   0.10 & $\pm$0.11  & $-$0.52 & $\pm$0.13  &         & ...        &         &...        \\
F22058$-$3501 & 1302300 &   ...  &    19 & 13 & 34  &    407&$\pm$20  &   0.97 & $\pm$0.17  &    0.09 & $\pm$0.33  &         & ...        &         &...        \\
F22206$-$2715 & 2203300 &   ...  &    43 &  6 & 97  &    217&$\pm$76  &$-$0.06 & $\pm$0.19  & $-$0.36 & $\pm$0.05  &         & ...        &         &...        \\
F22560$-$3501 & 2202322 &   ...  &    51 &  3 & 78  &    274&$\pm$66  &   0.15 & $\pm$0.32  & $-$0.42 & $\pm$0.29  & $-$0.48 & $\pm$0.28  &         &...        \\
F23128$-$5919 & 3302222 &   ...  &    52 &  2 & 57  &    360&$\pm$19  &   0.33 & $\pm$0.17  & $-$0.69 & $\pm$0.41  & $-$0.60 & $\pm$0.31  &         &...        \\
F23253$-$5415 & 3223300 &   ...  &    40 &  2 & 04  &    733&$\pm$99  &   0.07 & $\pm$0.30  & $-$0.01 & $\pm$0.06  &         & ...        & $-$0.79 &$\pm$0.23  \\
 09022$-$3615 & 3333333 &  154.7 &   164 &  5 & 67  &    547&$\pm$11  &   0.06 & $\pm$0.04  & $-$0.48 & $\pm$0.16  & $-$0.58 & $\pm$0.16  & $-$1.43 &$\pm$0.19  \\
\hline
\end{tabular}
\begin{flushleft}
Column descriptions: (1) Object name in the {\it IRAS} catalogue.
(2) Line flux qualities at \Hb, \OIII$\lambda5007$, \OI$\lambda6300$, \Ha, \NII$\lambda6584$, \SII$\lambda6717$, and \SII$\lambda6731$
(0=unmeasurable, 1=low, 2=moderate, 3=high).
(3) Absolute flux of \Ha\ [10$^{-15}$ ergs s$^{-1}$ cm$^{-2}$]. The values from uncalibrated spectra are not presented.
(4) Equivalent width of \Ha\  [\AA].
(5) Observed \Ha/\Hb\ line ratio. %The values from 2dF and 6dF spectra are not presented.
(6) \OIII\  line width after instrumental resolution correction and its uncertainty [km s$^{-1}$].
Line widths below 100 km s$^{-1}$ are not shown.
(7-10) Logarithms of the line ratios and their uncertainties. 
The ratios are corrected for Galactic extinction and stellar absorption.
The internal extinction correction is applied, if available.
%The large uncertainties ($>$ 0.2 dex) are flagged with a colon.
\end{flushleft}
\end{center}
\end{table*}

\begin{table*}
\begin{flushleft}
\textbf{Table 3.} Line information of the Sample B ULIRGs.\\
\end{flushleft}
\begin{center}
\begin{tabular}{rcr@{}lr@{}lc|crcr@{}lr@{}l}
\hline
\multicolumn{1}{c}{{\it IRAS} name} & quality &
\multicolumn{2}{c}{FWHM$_{\scriptsize \OIII}$}    &
\multicolumn{2}{c}{~$\log \dfrac{\OIII}{\Hb}$} & & &
\multicolumn{1}{c}{{\it IRAS} name} & quality &
\multicolumn{2}{c}{FWHM$_{\scriptsize \OIII}$}    &
\multicolumn{2}{c}{~$\log \dfrac{\OIII}{\Hb}$}\\
\hline
F00050$-$3259 & 33 &   378&$\pm$41  & $-$0.11&$\pm$0.14 & & &  F12432$-$3138 & 12 &   701&$\pm$66  &    0.74&$\pm$0.39\\
F00091$-$3905 & 12 &   621&$\pm$102 &    0.78&$\pm$0.40 & & &  F12452$-$2032 & 33 &   185&$\pm$53  &    0.56&$\pm$0.13\\
F00184$-$3331 & 11 &   472&$\pm$141 & $-$0.43&$\pm$0.38 & & &  F13269$-$2251 & 11 &   384&$\pm$93  & $-$0.05&$\pm$0.51\\
F00318$-$3137 & 22 &      &...      &    0.26&$\pm$0.30 & & &  F13270$-$0331 & 33 &  1146&$\pm$8   &    0.93&$\pm$0.14\\
F00569$-$3108 & 12 &      &...      &    0.84&$\pm$0.52 & & &  F13306$-$1644 & 12 &   175&$\pm$48  &    0.53&$\pm$0.34\\
F01004$-$2237 & 32 &   850&$\pm$28  &    0.55&$\pm$0.16 & & &  F13379$-$0256 & 23 &   310&$\pm$58  &    0.02&$\pm$0.16\\
F01009$-$3241 & 33 &   786&$\pm$13  &    1.09&$\pm$0.07 & & &  F13531$-$3422 & 23 &   760&$\pm$23  &    0.90&$\pm$0.19\\
F01160$-$2551 & 33 &   250&$\pm$54  & $-$0.15&$\pm$0.09 & & &  F14021$-$3139 & 12 &   575&$\pm$87  &    0.31&$\pm$0.37\\
F01212$-$5025 & 22 &   908&$\pm$134 &    0.13&$\pm$0.39 & & &  F14207$-$2002 & 12 &   448&$\pm$91  &    0.56&$\pm$0.53\\
F01234$-$0447 & 22 &      &...      & $-$0.17&$\pm$0.20 & & &  F14248$-$3644 & 12 &   331&$\pm$90  &    0.34&$\pm$0.41\\
F01358$-$3300 & 22 &   368&$\pm$18  &    0.25&$\pm$0.09 & & &  F14254$-$2655 & 11 &   711&$\pm$61  &    0.62&$\pm$0.45\\
F01379$-$3203 & 23 &  1176&$\pm$27  &    0.77&$\pm$0.19 & & &  F15130$-$1958 & 12 &  1283&$\pm$75  &    0.72&$\pm$0.47\\
F02038$-$0816 & 13 &   277&$\pm$28  &    0.94&$\pm$0.38 & & &  F16159$-$0402 & 11 &   833&$\pm$134 & $-$0.07&$\pm$0.43\\
F02068$-$2000 & 22 &      &...      &    0.52&$\pm$0.25 & & &  F19548$-$6237 & 11 &   335&$\pm$119 & $-$0.18&$\pm$0.51\\
F02130$-$1948 & 12 &   657&$\pm$83  &    0.27&$\pm$0.29 & & &  F20023$-$5253 & 12 &   376&$\pm$135 &    0.03&$\pm$0.35\\
F02356$-$4628 & 33 &   569&$\pm$45  & $-$0.13&$\pm$0.15 & & &  F20066$-$1630 & 32 &   134&$\pm$102 & $-$0.28&$\pm$0.30\\
F02364$-$4751 & 33 &   508&$\pm$40  &    0.43&$\pm$0.10 & & &  F20181$-$2244 & 33 &   882&$\pm$12  &    0.56&$\pm$0.12\\
F02384$-$1744 & 12 &   248&$\pm$66  & $-$0.07&$\pm$0.43 & & &  F20248$-$3204 & 12 &   359&$\pm$120 &    0.29&$\pm$0.54\\
F03000$-$2719 & 23 &   664&$\pm$17  &    0.92&$\pm$0.02 & & &  F20270$-$4237 & 12 &   696&$\pm$78  &    1.13&$\pm$0.37\\
F03130$-$3119 & 11 &   281&$\pm$83  &    0.22&$\pm$0.47 & & &  F20273$-$6558 & 13 &  1402&$\pm$70  &    1.07&$\pm$0.30\\
F03259$-$3105 & 31 &   256&$\pm$36  &    0.01&$\pm$0.32 & & &  F20542$-$1832 & 23 &   418&$\pm$24  &    0.58&$\pm$0.22\\
F03485$-$1827 & 33 &   657&$\pm$11  &    0.92&$\pm$0.12 & & &  F21016$-$1900 & 12 &   863&$\pm$165 & $-$0.03&$\pm$0.31\\
F03569$-$2535 & 23 &   157&$\pm$37  &    0.17&$\pm$0.19 & & &  F21356$-$1015 & 23 &   902&$\pm$12  &    0.50&$\pm$0.20\\
F04056$-$5722 & 32 &      &...      &    0.27&$\pm$0.18 & & &  F21435$-$3648 & 13 &   557&$\pm$25  &    0.76&$\pm$0.18\\
F04279$-$3035 & 23 &   116&$\pm$37  &    0.36&$\pm$0.20 & & &  F21488$-$2819 & 33 &   346&$\pm$13  &    0.33&$\pm$0.09\\
F04489$-$1026 & 13 &   525&$\pm$53  &    0.44&$\pm$0.19 & & &  F21555$-$4235 & 33 &   123&$\pm$32  &    0.09&$\pm$0.14\\
F05020$-$2941 & 33 &   154&$\pm$46  &    0.01&$\pm$0.15 & & &  F22301$-$2822 & 32 &  1083&$\pm$113 &    0.02&$\pm$0.19\\
F05156$-$3024 & 23 &   990&$\pm$28  &    0.76&$\pm$0.17 & & &  F22521$-$3929 & 13 &   775&$\pm$79  &    0.31&$\pm$0.40\\
F06158$-$4017 & 22 &   767&$\pm$52  &    0.28&$\pm$0.28 & & &  F22546$-$2637 & 23 &   194&$\pm$37  &    0.10&$\pm$0.16\\
F08411$-$2501 & 31 &   908&$\pm$149 &    0.46&$\pm$0.41 & & &  F23046$-$3454 & 33 &   309&$\pm$23  &    0.17&$\pm$0.10\\
F09090$-$1349 & 12 &      &...      & $-$0.07&$\pm$0.52 & & &  F23142$-$0611 & 22 &   452&$\pm$51  &    0.50&$\pm$0.38\\
F09521$-$0400 & 12 &   845&$\pm$94  &    0.22&$\pm$0.40 & & &  F23185$-$0328 & 33 &   956&$\pm$32  &    0.42&$\pm$0.15\\
F10298$-$2300 & 23 &  1490&$\pm$71  &    0.99&$\pm$0.24 & & &  F23206$-$1222 & 33 &   451&$\pm$5   &    0.46&$\pm$0.15\\
F10549$-$3702 & 12 &   491&$\pm$112 &    0.22&$\pm$0.28 & & &  F23242$-$0357 & 22 &      &...      &    0.20&$\pm$0.38\\
F11053$-$2413 & 13 &   521&$\pm$55  &    0.89&$\pm$0.44 & & &  F23516$-$2420 & 21 &      &...      & $-$0.39&$\pm$0.43\\
F11204$-$2154 & 22 &   391&$\pm$87  &    0.02&$\pm$0.36 & & &  F23529$-$2119 & 12 &   634&$\pm$106 &    0.50&$\pm$0.51\\
F11300$-$0522 & 32 &      &...      & $-$0.45&$\pm$0.26 & & &  F23559$-$3009 & 11 &   485&$\pm$71  & $-$0.19&$\pm$0.54\\
F12131$-$2809 & 23 &   713&$\pm$40  &    0.47&$\pm$0.17 & & &                &    &      &         &        &         \\
\hline
\end{tabular}
\begin{flushleft}
Column descriptions: (1, 5) Object name in the {\it IRAS} catalogue.
(2, 6) Line flux qualities at \Hb\ and \OIII$\lambda5007$ (0=unmeasurable, 1=low, 2=moderate, 3=high).
(3, 7) \OIII\  line width after instrumental resolution correction and its uncertainty [km s$^{-1}$].
Line widths below 100 km s$^{-1}$ are not presented.
(4, 8) Logarithm of \OIII/\Hb\ line ratio and its uncertainty. 
The ratio is corrected only for Galactic extinction and stellar absorption.
%The large uncertainties ($>$ 0.2 dex) are flagged with a colon.
\end{flushleft}
\end{center}
\end{table*}

We use the emission lines with flux uncertainty $<$ 60\% for spectral classification.
The qualities of line fluxes with uncertainty $<$ 20\%, 20--40\%, and 40--60\% are referred to hereafter as high, moderate, and low,
respectively (see Tables 2 and 3).

In the sample of 115 ULIRGs, 8 ULIRGs have broad Balmer lines,
FWHM of broad component of Balmer lines $>$ 2000 km s$^{-1}$ 
and height of the broad component $>$ 3 times the local rms of the continuum-subtracted spectra.
These are considered to be broad-line AGN galaxies\footnote{In Table 1, we use the term ``Seyfert 1 galaxies" to indicate
broad-line AGNs, which is commonly used in previous studies.}.

There are 32 narrow emission-line ULIRGs for which more than two line ratios (\OIII$\lambda5007$/\Hb\ and 
at least one of \NII$\lambda6584$/\Ha, \SII$\lambda \lambda6717,6731$/\Ha, and \OI$\lambda6300$/\Ha) were measured (hereafter Sample A).
In Fig. 2, we show the diagnostic diagrams for these ULIRGs and divide them into star-forming, (starburst-AGN) composite,
and narrow-line AGN galaxies based on their loci in the diagrams.
Star-forming galaxies lie below the pure star formation line (Kauffmann et al. 2003) in the \NII/\Ha\ diagram and 
lie below the extreme starburst line (Kewley et al. 2001) in other diagrams.
Composite galaxies lie between the extreme starburst line and the pure star formation line in the \NII/\Ha\ diagram.
Narrow-line AGNs lie above the extreme starburst line in all three diagrams.
Whenever possible, narrow-line AGNs are subdivided into Seyfert 2 and low ionization narrow emission-line region (LINER) galaxies.
Seyfert 2 galaxies lie above the Seyfert-LINER classification lines (Kewley et al. 2006) in the \SII/\Ha\ and \OI/\Ha\ diagrams,
whereas LINERs lie below the lines.
For ambiguous galaxies that are classified as one type in two diagrams but another type in the remaining diagram,
we adopt the types that are given in the first two diagrams.

There are 75 narrow emission-line ULIRGs without measurable \Ha\ line mainly because \Ha\ falls outside the spectral coverage (hereafter Sample B).
These galaxies could not be classified in the diagnostic diagrams so that we attempt to classify them
in flux ratio between \OIII\ and \Hb\ lines versus \OIII\ line width diagram as demonstrated in Fig. 3.
Zakamska et al. (2003) used the criteria of $\log$~(\OIII/\Hb)~$>$~0.3 and FWHM$_{\scriptsize \OIII}$~$>$~400 km s$^{-1}$
to select narrow-line AGN galaxies (dashed line).
Using our Sample A and two large, homogeneous samples in the literature
(the {\it IRAS} 1 Jy ULIRGs: Kim \& Sanders 1998; Veilleux et al. 1999;
Yuan et al. 2010; the SDSS ULIRGs: Hou et al. 2009),
we determine a new boundary to separate narrow-line AGN galaxies from the others (solid line)
with high completeness and reliability as far as possible:
$\log$~(\OIII/\Hb)~$>$~$-$0.13~(FWHM$_{\scriptsize \OIII}$/100~km~s$^{-1}$)~$+$~0.85.
If we regard AGNs classified in the diagnostic diagrams as genuine AGNs,
the application of our new boundary provides 89\% completeness (among the total 88 AGNs, 78 AGNs are found inside this boundary) 
and 89\% reliability (among 88 objects in the boundary, 78 objects are AGNs),
while 68\% (60/88) completeness and 90\% (60/67) reliability using the criteria of Zakamska et al.
The galaxies outside the boundary are mostly star-forming or composite galaxies,
but they are not separable from each other in this diagram. 
They are referred to as non-AGN galaxies.

Our 115 southern ULIRGs contain 8 broad-line AGNs, 49 narrow-line AGNs including four Seyfert 2 and one LINER galaxies,
and 58 non-AGNs including thirteen composite and twelve star-forming galaxies.
The spectral classification results (and their basic information) are presented in Table 1.
Detailed line information of Samples A and B is listed in Tables 2 and 3, respectively.

\section{Discussion}

\subsection{Reliability of our spectral classification}

\begin{table*}
\begin{flushleft}
\textbf{Table 4.} AGN fraction as a function of infrared properties \\
\end{flushleft}
%\vskip 3.0mm
\begin{center}
\begin{tabular}{lcccc}
\hline
 & \multicolumn{4}{c}{$\log~(L_{\mbox{\scriptsize IR}}/L_\odot)$}\\[3pt]
\cline{2-5}\\[-7pt]
\multicolumn{1}{c}{sample} &
\multicolumn{1}{c}{all} &
\multicolumn{1}{c}{[12.0, 12.15)} &
\multicolumn{1}{c}{[12.15, 12.4)} &
\multicolumn{1}{c}{[12.4, 13.0)}\\
\hline
This study            & 50\% of 115  & 48\% of 61  & 50\% of 40  & 57\% of 14\\
Kim \& Sanders (1998) & 45\% of 108  & 38\% of 40  & 39\% of 49  & 79\% of 19\\
Hou et al. (2009)     & 55\% of 209  & 43\% of 90  & 51\% of 81  & 89\% of 36\\
\hline
 & \multicolumn{4}{c}{$\log~(f_{25}/f_{60})$}\\[3pt]
\cline{2-5}\\[-7pt]
\multicolumn{1}{c}{sample} &
\multicolumn{1}{c}{all} &
\multicolumn{1}{c}{[$-$1.4, $-$0.9)} &
\multicolumn{1}{c}{[$-$0.9, $-$0.5)} &
\multicolumn{1}{c}{[$-$0.5, $-$0.1)}\\
\hline
This study            & 59\% of  37  & 40\% of 10  & 47\% of 15  & ~92\% of 12\\
Kim \& Sanders (1998) & 56\% of  59  & 29\% of 28  & 76\% of 25  & 100\% of ~6\\
Hou et al. (2009)     & 53\% of  45  & 33\% of 18  & 38\% of 13  & ~93\% of 14\\
\hline
 & \multicolumn{4}{c}{$\log~(f_{60}/f_{100})$}\\[3pt]
\cline{2-5}\\[-7pt]
\multicolumn{1}{c}{sample} &
\multicolumn{1}{c}{all} &
\multicolumn{1}{c}{[$-$0.7, $-$0.3)} &
\multicolumn{1}{c}{[$-$0.3, $-$0.1)} &
\multicolumn{1}{c}{[$-$0.1,    0.3)}\\
\hline
This study            & 40\% of  ~70  & 36\% of 11  & 33\% of 36  & 52\% of 23\\
Kim \& Sanders (1998) & 45\% of  108  &             & 33\% of 51  & 56\% of 57\\
Hou et al. (2009)     & 43\% of  106  & 58\% of 19  & 39\% of 61  & 46\% of 24\\
\hline
\end{tabular}
\end{center}
\end{table*}

The small aperture spectroscopy could not always contain enough light of an extended source to determine its spectral type.
For reliable classification, it is suggested to use an aperture covering more than $\sim$ 20\% of the galaxy light.
The minimum aperture covering fraction of 20\% corresponds to 2.1 kpc (see fig. 6 in Kewley et al. 2005).
All spectra of ULIRGs in this study satisfy this condition.
Therefore, aperture-related effects on our classification are expected to be negligible.

In the 2dFGRS and 6dFGS spectra, the flux for individual line can be unreliable because these spectra are not properly flux-calibrated.
Nevertheless, the flux ratio between lines with similar wavelengths is still useful (e.g., Mouhcine et al. 2005; Owers et al. 2007).
To ensure this, in Fig. 4,
we compare \OIII/\Hb\ line ratios derived from both calibrated (i.e., CTIO or SDSS) and uncalibrated (i.e., 2dFGRS or 6dFGS) spectra.
It shows that two measurements agree well within the errors.
The measurements between the 2dFGRS and 6dFGS spectra also agree well.

Among the 115 ULIRGs, there are 22 objects whose spectral types have been previously determined in the literature.
The spectral types for 19 out of 22 ULIRGs determined in this study are consistent with those in the literature.
If we compare ULIRGs whose subclasses (see columns 11 and 12 in Table 1) were determined, 9 out of 13 ULIRGs show a good agreement.
Although there are three ULIRGs with different types (or four ULIRGs with different subclasses),
all of these ULIRGs are composite galaxies either in this study or in the previous studies.

\begin{figure}
	\begin{center}
\includegraphics [width=90mm] {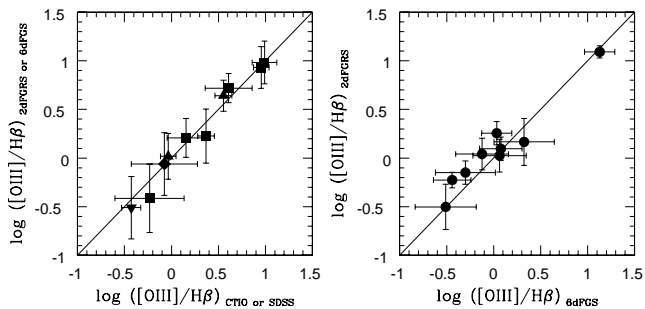}
\end{center}
\caption{Comparison of \OIII/\Hb\ line ratio between two spectra.
In the left panel, the measurements from CTIO-6dFGS, CTIO-2dFGRS, SDSS-6dFGS, and SDSS-2dFGRS
are denoted by squares, diamonds, triangles, and inverse triangles, respectively.
In the right panel, the measurements from 6dFGS-2dFGRS are denoted by circles.
Only objects with S/Ns of both spectra $>$ 3 are presented.
The one-to-one relation (solid line) is overplotted.}
\end{figure}

%\clearpage

\subsection{Dependence of optical properties on infrared parameters}

In our sample, 50\% (57/115) of the ULIRGs are found to host AGN.
The AGN fraction depends on the infrared luminosity and the {\it IRAS} flux density ratios (hereafter {\it IRAS} colours) as follows:
48\% for ULIRGs with 12.0 $\leqslant$ $\log~(L_{\mbox{\scriptsize IR}}/L_\odot)$ $<$ 12.15, 
50\% for [12.15, 12.4), and 57\% for [12.4, 13.0).
40\% for ULIRGs with $-$1.4 $\leqslant$ $\log~(f_{25}/f_{60})$ $<$ $-$0.9,
47\% for [$-$0.9, $-$0.5), and 92\% for [$-$0.5, $-$0.1).
36\% for ULIRGs with $-$0.7 $\leqslant$ $\log~(f_{60}/f_{100})$ $<$ $-$0.3,
33\% for [$-$0.3, $-$0.1), and 52\% for [$-$0.1, 0.3).
The ULIRGs with a flux upper limit at 25 (100) $\mu$m are not included in the calculations for {\it IRAS} 25$-$60 (60$-$100) $\mu$m colour dependence.
These results are summarized in Table 4 together with those from the 1 Jy (Kim \& sanders 1998) and SDSS (Hou et al. 2009) samples.
Note that no ULIRGs in the 1 Jy sample have colours of $\log~(f_{60}/f_{100})$ $<$ $-$0.3 due to the selection criteria in Kim \& Sanders (1998).
The AGN fractions for each sample are comparable in the sense that the differences are less than 1.6 $\sigma$ by assuming Poisson errors.
In all samples, there is a tendency for the ULIRGs with higher infrared luminosity, warmer {\it IRAS} 25$-$60 $\mu$m colour to be more AGN-like.
These findings are consistent with those in previous works 
(e.g., Veilleux et al. 1999; Kewley et al. 2001; Goto 2005; Cao et al. 2006; Hou et al. 2009; Yuan et al. 2010).
%For our sample, we use $L_{\mbox{\scriptsize IR}}$ mainly in Hwang et al. (2007), 
%which were calculated from spectral energy distribution fitting method with the M82 template.
%The results do not change when we use $L_{\mbox{\scriptsize IR}}$ based on {\it IRAS} 60 and 100 $\mu$m flux densities as in the other studies.

\begin{figure*}
\begin{center}
\includegraphics [width=140mm] {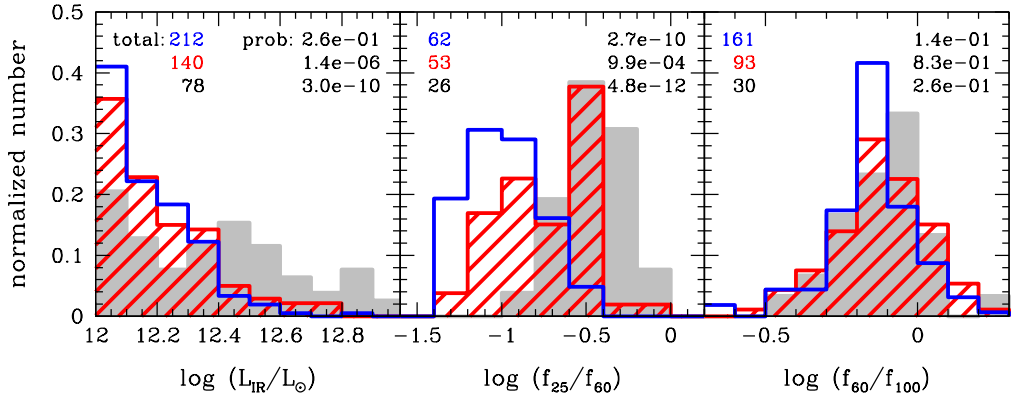}
\end{center} 
\caption{Infrared luminosity (left), {\it IRAS} 25$-$60 $\mu$m colour (middle), and {\it IRAS} 60$-$100 $\mu$m colour (right) distributions
of the combined ULIRG sample. The distributions of non-AGN, narrow-line AGN, and broad-line AGN ULIRGs are demonstrated
by solid, hatched, and shaded histograms, respectively. %$\setminus$
The ULIRGs with a flux upper limit at 25 (100) $\mu$m are not included in the middle (right) panel.
The histograms are normalized to the total numbers of each spectral type, which are represented in the upper-left corner
(upper: non-AGN; middle: narrow-line AGN; lower: broad-line AGN).
%The values in the upper-right corner show the maximum deviation between cumulative distribution functions of two types of ULIRGs, 
%given by the Kolmogorov-Smirnov test (Large values mean that the two distributions are significantly different.).
The values in the upper-right corner show the probability that the two types of ULIRGs are drawn from the same population, 
given by the K-S test (Small values mean that the distributions of two types are significantly different.).
The upper, middle, and lower values are calculated between non-AGNs and narrow-line AGNs, narrow-line AGNs and broad-line AGNs, and
broad-line AGNs and non-AGNs, respectively.}
\end{figure*}

\begin{figure*}
\begin{center}
\includegraphics [width=150mm] {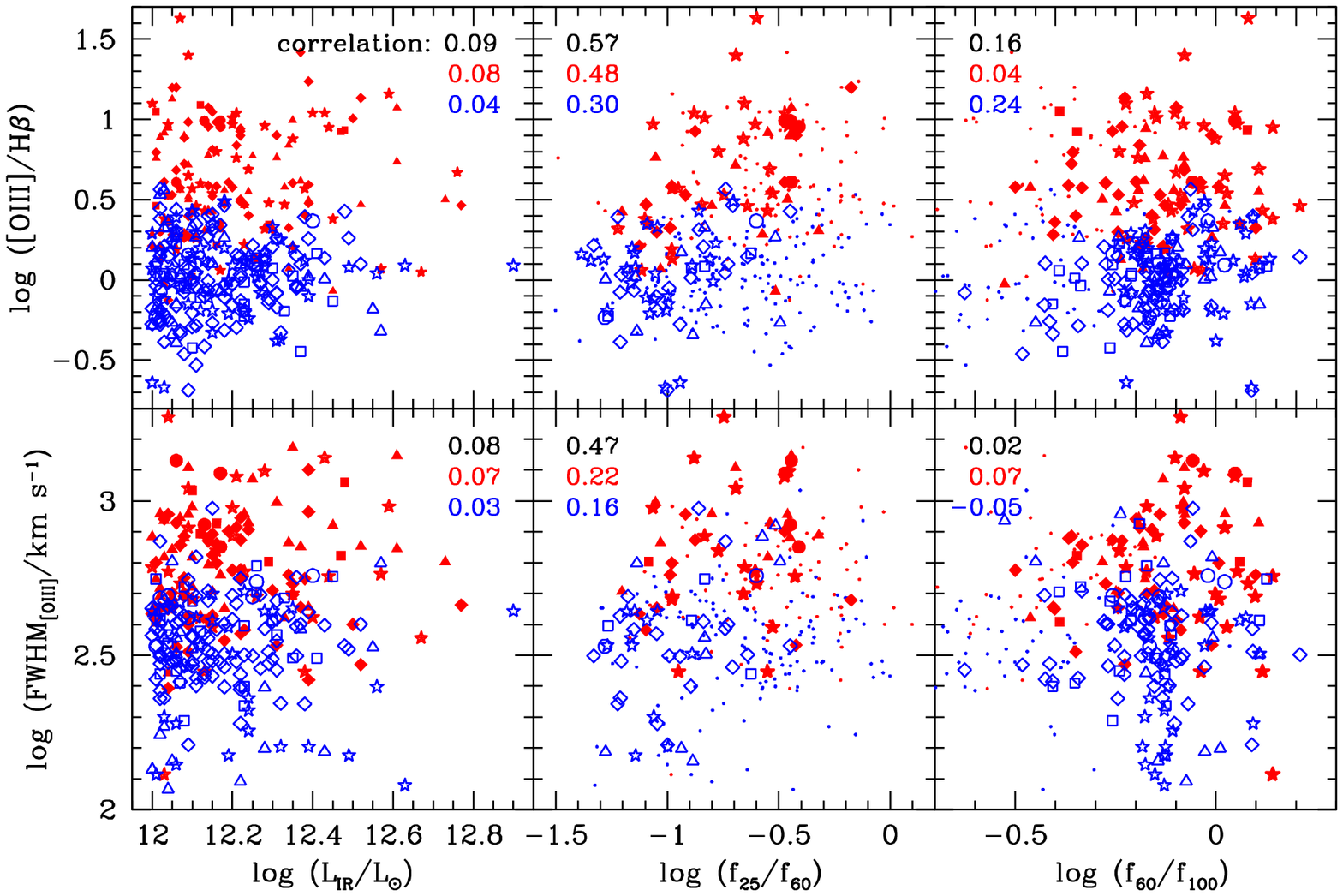}
\end{center}
\caption{Relations between optical and infrared properties of narrow emission-line objects in the combined ULIRG sample.
\OIII/\Hb\ line ratio vs. infrared luminosity (upper-left),
\OIII/\Hb\ line ratio vs. the {\it IRAS} 25$-$60 $\mu$m colour (upper-middle),
\OIII/\Hb\ line ratio vs. the {\it IRAS} 60$-$100 $\mu$m colour (upper-right),
FWHM$_{\scriptsize \OIII}$ vs. infrared luminosity (lower-left),
FWHM$_{\scriptsize \OIII}$ vs. the {\it IRAS} 25$-$60 $\mu$m colour (lower-middle) and
FWHM$_{\scriptsize \OIII}$ vs. the {\it IRAS} 60$-$100 $\mu$m colour (lower-right)
diagrams are presented. The symbol shapes are the same as in Fig. 3.
The filled and open symbols are narrow-line AGNs and non-AGNs, respectively.
The values in each panel indicate Spearman's rank correlation coefficient between two properties.
The upper, middle, and lower values are calculated from narrow-line AGN plus non-AGN, narrow-line AGN, and non-AGN ULIRGs, respectively.
In the middle (right) panels, the ULIRGs which have a flux upper limit at 25 (100) $\mu$m are not used in the calculations
and they are represented by dots.}
\end{figure*}

In Fig. 5, we present infrared luminosity, {\it IRAS} 25$-$60 $\mu$m colour, and {\it IRAS} 60$-$100 $\mu$m colour distributions of
the combined sample (432 ULIRGs) including this study, Kim \& Sanders (1998), and Hou et al. (2009).
Based on the Kolmogorov-Smirnov (K-S) test, 
we find that the infrared luminosity distribution of broad-line AGN ULIRGs is significantly different from those of non-AGN and narrow-line AGN ULIRGs,
while those of non-AGN and narrow-line AGN ULIRGs are not so different. 
We also find that the three types of ULIRGs are significantly different from each other in {\it IRAS} 25$-$60 $\mu$m colour distribution but
they are indistinguishable in {\it IRAS} 60$-$100 $\mu$m colour distribution.
In {\it IRAS} 25$-$60 $\mu$m colour distribution, the significant difference between AGN and non-AGN ULIRGs strengthens
that mid-infrared colour is a good indicator of AGN activity in infrared galaxies (e.g., de Grijp et al. 1985; Neff \& Hutchings 1992).
The little differences in {\it IRAS} 60$-$100 $\mu$m colour distribution reinforce that
star formation dominates the emission of AGN in the far-infrared regime (e.g., Elbaz et al. 2010; Hatziminaoglou et al. 2010; Hwang et al. 2010b; Shao et al. 2010).
On the other hand, narrow-line AGN ULIRGs have in general lower infrared luminosity and cooler {\it IRAS} 25$-$60 $\mu$m colour than broad-line AGN ULIRGs do.
If we assume that the infrared emission in galaxies is emitted isotropically, and does not depend on viewing angle 
(e.g., Mulchaey et al. 1994; Schartmann et al. 2008; Gandhi et al. 2009),
the difference between narrow-line AGN and broad-line AGN ULIRGs seems not to be compatible
with the predictions of the orientation-dependent unification model of AGNs
in which narrow-line and broad-line AGNs are intrinsically same objects observed from different angles (Antonucci 1993). 
This conflict can be explained if narrow-line AGN ULIRGs host a central engine deeply buried by extended, dusty regions of star formation 
as proposed by several authors (e.g., Genzel et al 1998; Gerssen et al. 2004).

In Fig. 6, we show the relations between optical (\OIII/\Hb\ line ratio or \OIII\ line width) and infrared
(infrared luminosity, {\it IRAS} 25$-$60 $\mu$m colour or {\it IRAS} 60$-$100 $\mu$m colour) properties of the combined ULIRG sample.
Broad-line AGN ULIRGs are not plotted because their line widths and fluxes are not properly measured in many cases\footnote{
The spectra of broad-line AGNs should be analysed with more sophisticated methods of galaxy modelling (e.g., Kim et al. 2006).}.
Since the spectral types are determined by the optical emission lines, narrow-line AGNs and non-AGNs are well separated along the y-axis.
As expected, clear correlations are found only in the middle panels (see the values of Spearman's rank correlation coefficients).
We checked these relations using various infrared colours from the {\it AKARI} all-sky survey point source catalogues\footnote{
http://www.ir.isas.jaxa.jp/AKARI/Observation/PSC/Public/} 
but we could not draw any meaningful results due to the small number of data points.

\section{Summary}
We studied optical spectral properties of 115 southern ULIRGs using the spectra 
obtained from our CTIO observations, 2dFGRS, and 6dFGS.
For ULIRGs with \Ha\ measurement, we classified them in the standard diagnostic diagrams.
We classified the other ULIRGs using the \OIII$\lambda5007$/H$\beta$ line ratio against \OIII$\lambda5007$ line width diagram
with an empirically determined criterion.
Main results are summarized below.
\begin{enumerate}
\item Our new criterion, $\log$~(\OIII/\Hb)~$>$~$-$0.13~(FWHM$_{\scriptsize \OIII}$ /100~km~s$^{-1}$)~$+$~0.85, is successful 
to separate AGN ULIRGs from non-AGN ULIRGs with completeness and reliability of about 90\%.
\item In our sample of the 115 ULIRGs, there are 8 broad-line AGNs, 49 narrow-line AGNs, and 58 non-AGNs.
The AGN fraction is 50\% and changes as a function of infrared luminosity and {\it IRAS} 25$-$60 $\mu$m colour.
These results are consistent with those in previous studies.
\item Using the combined ULIRG sample, we show that the colour distributions of AGN and non-AGN ULIRGs are significantly different 
in {\it IRAS} 25$-$60 $\mu$m colour and are indistinguishable in {\it IRAS} 60$-$100  $\mu$m colour.
These results support that mid-infrared colour is sensitive to AGN activity, whereas far-infrared colour is dominated by star formation.
\item We also show that broad-line AGN ULIRGs differ from narrow-line AGN ULIRGs in infrared luminosity and {\it IRAS} 25$-$60 $\mu$m colour.
This presents a challenge to the simple unification model of AGNs.
\end{enumerate}

\section*{Acknowledgments}
We thank the CTIO staff for their help on the observations.
We also thank L. G. Hou for providing the \OIII\ line width data of the SDSS ULIRGs.
We are grateful to anonymous referee whose comments helped to improve the original manuscript.
This work was supported by Mid-career Research Program through NRF grant 
funded by the MEST (No.2010-0013875).
H.S.H acknowledges the support of the Centre National d'Etudes Spatiales (CNES).
S.C.K is a member of the Dedicated Researchers for Extragalactic AstronoMy (DREAM) team
in Korea Astronomy and Space Science Institute (KASI).

\label{lastpage}

\end{document}